\documentclass[a4paper]{article}   
\usepackage{a4wide}

\usepackage{mathrsfs}
\usepackage{latexsym,pifont,color}
\usepackage[english]{babel}
\usepackage{calrsfs}
\usepackage{stmaryrd}
\usepackage{amsmath,amssymb}
\usepackage{empheq,soul}
\usepackage{color}
\usepackage{upgreek}
\usepackage{bbold}
\usepackage{hyperref}
\usepackage{etoolbox}
\usepackage{xspace}
\usepackage{tikz}
\usepackage{proof}
\usepackage{flushend}

\newcommand{\sv}[1]{{\small \texttt{#1}}}

\usepackage[final]{listings}
\lstdefinestyle{INLINE}{
  basicstyle=\small\tt,
  numbers=left,
  numberstyle=\scriptsize\color{red},
  numbersep=1em,
}
\lstdefinestyle{DISPLAY}{
}
\lstdefinestyle{FLOAT}{
  float,
  captionpos=b,
  basicstyle=\small\tt,
  numbers=left,
  numberstyle=\scriptsize\color{red},
  numbersep=1em,
}
\lstdefinelanguage{ObjectiveJoinCalculus}{
  morekeywords={null,let,def,obj,or,in,null,if,then,else,mod},
  basicstyle=\tt,
  sensitive=true,
  keywordstyle=\color{blue},
  commentstyle=\color{magenta},
  columns=fixed,
  mathescape=true,
  basewidth=0.5em,
  comment=[s]{(*}{*)},
  literate={=>}{$\triangleright$}1
           {<-}{$\leftarrow$}1
}

\lstdefinelanguage{MYSCALA}{
  morekeywords={def,val,new,yield,class,extends,<-,=>,case,with,type,object,trait,private,class,return,abstract,false,true,if,else,for},
  sensitive=true,
  keywordstyle=\color{blue},
  commentstyle=\color{magenta},
  showstringspaces=false,
  columns=fixed,
  mathescape=true,
  basewidth=0.5em,
  comment=[l]{//},
  literate={and1}{and}3
           {and2}{and}3
           {and3}{and}3
           {and4}{and}3,
}

\lstnewenvironment{SF}[1][]{\lstset{language=MYSCALA,style=FLOAT,#1}}{}
\lstnewenvironment{SI}[1][]{\lstset{language=MYSCALA,style=INLINE,#1}}{}

\begin{document}
\title{On the chemistry of typestate-oriented actors}
\author{Silvia Crafa \hspace{3cm} Luca Padovani \\
Universit\`a di Padova, Italy \hspace{1cm} Universit\`a di Torino, Italy}
\date{}
\maketitle

\begin{abstract}
  Typestate-oriented programming is an extension of the OO paradigm in
  which objects are modeled not just in terms of interfaces but also
  in terms of their usage protocols, describing legal sequences of
  method calls, possibly depending on the object's internal state. We
  argue that the Actor Model allows typestate-OOP in an inherently
  distributed setting, whereby objects/actors can be accessed
  concurrently by several processes, and local entities cooperate to
  carry out a communication protocol. In this article we illustrate
  the approach by means of a number of examples written in Scala
  Akka. We show that Scala's abstractions support clean and natural
  typestate-oriented actor programming with the usual asynchronous and
  non-blocking semantics. We also show that the standard type system
  of Scala and a typed wrapping of usual (untyped) Akka's
  \sv{ActorRef} are enough to provide rich forms of type safety so
  that well-typed actors respect their intended communication
  protocols.

This approach draws on a solid theoretical background, consisting of a
sound behavioral type system for the Join Calculus, that is a
foundational calculus of distributed asynchronous processes whose
semantics is based on the Chemical Abstract Machine, that unveiled its
strong connections with typestate-oriented programming of both
concurrent objects and actors. 
\end{abstract}

\section{Introduction}
A key issue of distributed systems is the coordination of distributed
entities that concurrently run in the system.  Very different
solutions can be developed to address this problem, but even
distributed systems can be very different, hence it is important to
carefully choose a solution (or a set of solutions) that is
appropriate for the specific distributed system under consideration.
A major distinction, indeed, can be made between the systems that
adopt the shared memory model and those that rely on message passing.
The first kind of systems is often a natural model for
\emph{data-centric} applications, that fit a \emph{centralized
  control} of distributed entities that cooperate to fulfill logically
centralized algorithms operating on mutable shared data.  On the other
hand, \emph{communication-centric} applications are usually built on
top of concurrent/distributes entities that take their decisions only
based upon local information (\emph{locality and isolation
  principles}).

A general approach to structuring distributed systems is
\emph{Protocol-Oriented Programming}, which endorses thinking (hence
programming) in terms of communication protocols. In this view,
programming a concurrent/distributed system entails the design of a
precise communication protocol involving a possibly dynamic set of
interacting parties. However, in order for a paradigm to become an
effective programming style, it requires high-level support to
\emph{express coordination protocols} in the programming language,
and, above all, it requires some support to \emph{check protocol
  compliance}. Finding the correct abstractions to express and enforce
coordination protocols is very hard, especially because suitable
abstractions should also productively interoperate with the other
abstractions provided by the language.  For instance, systems with
centralized control best fit a top-down implementation of a global
coordination protocol, e.g. the multiparty session types's methodology
that projects a global protocol into the sub-protocols of local
parties~\cite{MultipartySessions08}.  On the contrary, distributed
systems made of local entities that work with strong isolation and
locality principles, require support for consistent bottom-up
compositions of local behaviors.  Furthermore, real systems usually
rely on hybrid architectures and logical models, so it is important to
remind that there is no single solution and trade-offs are
unavoidable.

In this article we study the application of the Typestate-Oriented 
Programming discipline to achieve protocol-oriented programming in an
Actor System. Typestate-oriented programming  (TSOP for
short)~\cite{DeLineFahndrich04,AldrichEtAl09,SunshineEtAl11,GarciaEtAl14}
is an extension of the OO
paradigm where objects are modeled not just in terms of interfaces
but in terms of their usage protocols, describing legal sequences of
method calls, possibly depending on the object’s internal state. 
In \cite{CrafaPadovani15} we showed how TSOP can be statically
enforced in the Objective Join Calculus, that is a calculus of
concurrent objects equipped with a semantics 
based on the chemical abstract machine \cite{FournetGonthier96}. More
precisely, we introduced a simple behavioral type theory that allows
to describe and enforce structured object protocols consisting of
possibilities, prohibitions and obligations. According to the chemical
metaphor, programs are modeled as chemical soups of molecules
(i.e., multisets of messages sent to objects) that encode both the
current state of the objects and the (pending) operations on them.
Moreover, reaction rules, corresponding to object's method
definitions, explicitly specify both the valid combinations of state
and operations as well as the state changes engendered by each
operation. 

In \cite{CrafaPadovaniJournal} we put forward a couple of examples
written with Scala Akka actors, showing that the actor programming
model~\cite{ActorsFirst,Agha:ActorsModel} bears strong similarities
with TSOP we realized in the Objective Join Calculus, at least at the
program level.  The formal development of a behavioral type system for
actors along the lines of \cite{CrafaPadovani15} is the subject of
current work.  Instead, in this article we further explore the
implementation of TSOP in the context of Scala Akka, and even if a
full implementation of the behavioral type discipline of
\cite{CrafaPadovani15} requires a compiler extension, we demonstrate
that exploiting the powerful typing support offered by Scala allows us
to go a long way into checking protocol compliance solely using
off-the-shelf development tools.

\subsection{Road map and summary of the contributions}

The implementation of typestate-oriented actors is entirely
illustrated by means of examples. We start by fully studying an object
with a very simple usage protocol, but that is enough to display
all the issues of the typestate discipline.
This main running example is presented in stages so to separate the
explanation of different ingredients, more precisely: 
\begin{itemize}
\item in section 2.1 we draw on plain Akka actors, that are
  essentially untyped, thus in order to address protocol violations
  they have to resort to defensive programming by explicitly dealing
  with unintended messages;
\item in section 2.2 we introduce the use of types to represent
  actor's interfaces and typed actor references to encapsulate
  references to stateful actors;
\item in section 2.3 we resort to the explicit continuation-passing
  programming style to keep track of the dynamic change of the actors
  state;  
\item in section 2.4 we refactor the code so to just implicitly deal
  with continuations. We define a suitable monad that allows 
  ($i$) to simplify the code of the user of a stateful actor,
        bringing better to light its logic,
  ($ii$) the compiler to check that a stateful actor will only
         receive intended messages at the intended state,
  ($iii$) and usage protocol to be completely encoded in terms of
          typed expressions; 
\item while in sections 2.3 and 2.4 we considered a stateful actor
  accessed by a single user, in section 2.5 we show that in order to
  scale the approach to the case of multiple concurrent users it is
  sufficient for the stateful actor to mix-in (a trait implementing)
  the semantics \`a la chemical abstract machine.
\end{itemize} 
The final section 3 displays our TSOP approach at work on a more
complex scenario. It shows that Scala Akka actors effectively support
the definition of stateful actors that can dynamically change
\emph{both their behavior and their interface}. 
Moreover, the Scala type system is expressive enough to let the
compiler check that a stateful actor will only receive intended
messages at the intended states. Besides the absence of wrong
messages, a full typing support for protocol-oriented programming
would also require to detect violations of protocol obligations, that
is to ensure that all the intended messages are eventually sent. This
property is captured by the behavioral typing studied in
\cite{CrafaPadovani15} but the required linear types are beyond the
expressive power of the current Scala type system. 

All the code shown in the article is compatible with Scala 2.11.6 and
Akka 2.3.4.

\section{The 1-place Buffer}
\label{sec:Buffer}

In this section we consider as a running example a buffer that can
contain at most one value of a generic type \sv{T}. 
The buffer provides just two operations:
\sv{insert(x:T)} and \sv{remove()}, however since the buffer cannot
hold more than a value, no two consecutive \sv{insert} operations can
be executed, and similarly no two consecutive \sv{remove} operations.
Such a usage constraint can be rephrased in typestate-oriented terms
as the following protocol:

\begin{quote}
{\bf Buffer protocol:} the buffer has two possible \emph{states},
\sv{EMPTY} and \sv{FULL}. When in state \sv{EMPTY} its
\emph{interface} only contains the operation \sv{insert(x:T)}, whose
execution moves the buffer to state \sv{FULL}. When the buffer is in
state \sv{FULL} its \emph{interface} only contains the operation
\sv{remove()}, whose execution moves it to state \sv{EMPTY}. The
initial state of the buffer is \sv{EMPTY}. 
\end{quote}

\subsection{Untyped Buffer Actor}

Moving from the buffer's protocol above to the definition of a
\sv{Buffer} actor is straightforward: simply observe that actors can
directly implement finite state machines (FSM) and the protocol above
can be easily expressed as a FSM with two states \sv{EMPTY} and
\sv{FULL} and two edges labeled \sv{insert(x:T)} and \sv{remove()}
connecting one state with the other.  We point out that we are not
necessarily advocating the use of FSMs to express objects protocols,
we use them here just as a mental bridge between 
protocol declarations and actor implementations.

The following code defines 
a generic Akka actor \sv{Buffer[T]} that is intended to handle only
the messages \sv{insert} and \sv{remove}, implemented as case classes.
In particular, the actor defines two possible behaviors that
correspond to the two possible buffer's states: \sv{EMPTY} and
\sv{FULL}. Each behavior is a partial function (of type \sv{Receive}
as expected for Akka actor's behaviors)
that only handles the single message that belongs to
the interface of the corresponding state. Moreover, the state change 
engendered by the operations is obtained by using the actor's ability
to change its current behavior by means of the \sv{become}
method. Finally, the definition of the \sv{receive} method corresponds
to setting the initial state of the buffer to \sv{EMPTY}.  

\begin{SI}
case class insert[T](value:T) 
case class remove() 

class Buffer[T] extends Actor {
  def EMPTY:Receive = {
      case insert(x:T) => context.become(FULL(x)) 
                          println("inserted "+x)
  }
  def FULL(x:T):Receive = {
      case remove() => context.become(EMPTY) 
                       println("removed "+x)                   
  }
  def receive = EMPTY
}
\end{SI}
\noindent
Even if the code above is typed, there is no type support to guarantee
that: 
\begin{enumerate}
\item no messages other than \sv{insert} and \sv{remove} are sent to
  the buffer. 
\item the \sv{insert(v)} message is sent only to an \sv{EMPTY} 
buffer and the \sv{remove()} message is sent only to a \sv{FULL}
buffer. Therefore, since each message makes the buffer switch state,
\sv{insert} and \sv{remove} must be alternated.
\end{enumerate}
Indeed, the following user code is well-typed even if it violates the
intended buffer protocol:

\begin{SI}
val s = ActorSystem()
val buffer = s.actorOf(Props(new Buffer[Int]),"buff")
val user = s.actorOf(Props(new Actor{
    buffer ! insert(4) 
    buffer ! remove()  
    buffer ! insert(10)  
    buffer ! insert(20)  // the msg is received and silently discarded
    buffer ! 4           // the msg is received and silently discarded
    def receive = PartialFunction.empty
  }))
\end{SI}

The \sv{insert(20)} message is indeed received by a buffer whose
current behavior is the \sv{FULL(10)} partial function, which is not
defined on the \sv{insert} case class. The default Akka's policy for
actors that receive a message that is not handled by their current
behavior is to wrap that message into an \sv{UnhandledMessage} object
and forward it to the actor system's event stream for
logging. Therefore the reception of the unintended (logically wrong)
messages \sv{insert(20)} and \sv{4} gets completely unnoticed by the
user actor.
By adopting a defensive programming strategy, we could enrich the
\sv{Buffer} class with an overriding of the \sv{unhandled(msg:Any)}
method inherited from the \sv{Actor} class so to explicitly manage
unintended messages. However, the aim of the TSOP approach is
opposite: instead of protecting the buffer, it checks the user code
and statically marks the sending of the wrong messages as untyped
actions.

\subsection{Adding typed references}

The first one of the two guarantees formulated above reminds the basic
type safety property of OOP: in well-typed programs all the methods
invoked on objects belong to the object's interface. In the context
of actors there is no method invocation, but a message is passed as
a parameter to the dispatching method, that in Akka is the method
\sv{def !(msg:Any):Unit} of the class \sv{ActorRef}. However, in this
scenario there is no way to link (and check) the actor's interface
(i.e. the set of allowed incoming messages) with the types of the
parameters of the dispatching method.

We then resort to the following solution: first of all we use the
nominal type system of Scala to represent actors interfaces, then 
we wrap actor references of type \sv{ActorRef} into instances of the 
\sv{TypedRef[T]} generic class which provides a typed-version of the 
dispatching method with signature
 \sv{def tyTell(msg:T):Unit}\footnote{The name
  stands for ``typed Tell'', since ``tell'' is the name of
  the Akka method identified by the symbol \sv{!}.}. 
In other terms, a wrapper of type \sv{TypedRef[T]} encapsulates a
reference to a stateful actor \emph{at state} \sv{T}, and provides a
method that allows for the static checking of the correctness of the
incoming messages. Observe that the \sv{TypedRef[-T]} class is
contravariant in its type parameter: indeed, let \sv{S} be a
sub-interface of \sv{T}, then an actor of type \sv{TypedRef[S]}
can be safely substituted by an actor of type \sv{TypedRef[T]} which
is able to handle a superset of messages.

The following code applies this approach to the running example:
let \sv{BufferInterf}, \sv{ProduceInt} and \sv{ConsumeInt} be 
a hierarchy of empty traits corresponding to the interfaces assumed by
the buffer in different states; their names remind that when in state
\sv{EMPTY}, resp. \sv{FULL}, the buffer allows the production,
resp. consumption, of an item. Accordingly, the case classes extend
the suitable trait so that \sv{BufferInterf} is the (super-)type of
the buffer's messages. Let also \sv{buffer} be a reference of type 
\sv{TypedRef[BufferInterf]} built around the \sv{ActorRef} spawned by
the actor system upon creation of the \sv{Buffer} actor (lines 15-16).
Then the (typed-)invocation \sv{buffer tyTell(msg)} will only
compile if \sv{msg} is one of the two intended messages \sv{insert} or
\sv{remove}.

\begin{SI}
trait BufferInterf 
trait ProduceInt extends BufferInterf 
trait ConsumeInt extends BufferInterf  

case class insert[T](value:T) extends ProduceInt
case class remove()  extends ConsumeInt

class TypedRef[-T](r:ActorRef) {
  def tyTell(msg:T) =  r ! msg  // add a layer of typing over usual send
}

class Buffer[T] extends Actor ... // defined exactly as in section 2.1

val s = ActorSystem()
val untypedBuffer = s.actorOf(Props(new Buffer[Int]),"buff")
val buffer = new TypedRef[BufferInterf](untypedBuffer)

val user = s.actorOf(Props(new Actor{
    buffer tyTell insert(4) 
    buffer tyTell remove()  
    //buffer tyTell 4   does not compile
    untypedBuffer ! 4         // compiles but the msg is silently discarded
    buffer tyTell insert(10)  
    buffer tyTell insert(20)  // compiles but the msg is silently discarded

    def receive = PartialFunction.empty
  }))
\end{SI}

In the user code above we see the type checking in action to prevent
unintended messages to be sent to a \emph{typed actor reference}:
line 21 does not compile since the invocation of the \sv{tyTell}
method of the \sv{TypedRef[BufferInterf]} class does not accept the
parameter \sv{4} which is not of type \sv{BufferInterf}. On the
other hand, nothing prevents to send the wrong message \sv{4} to the
buffer actor by using its \emph{untyped} reference as in line 22. This
kind of errors can be avoided by designing a proper encapsulation of
the type \sv{ActorRef} so that actor users can only handle typed actor
references. 

More importantly, line 24 above shows that the second guarantee
required by the buffer protocol, that is the alternation of
\sv{insert} and \sv{remove}, eludes the type checking: two consecutive
insertions are well typed even if the second one gets silently
discarded because it is received by a \sv{FULL} buffer. This is not
surprising since the information encoded in the types just
refer to the buffer interfaces, but nothing expresses the state-change
(possibly) provoked by the reception of a message.
Furthermore, the initial \sv{buffer} reference in line 16
could be better typed as an empty buffer,
i.e. \sv{TypedRef[ProduceInt]}, since it initially accepts only
\sv{insert} messages. However, with such a type, no \sv{remove}
message can ever be passed to the \sv{tyTell} method, even after 
an initial insertion that would have moved the buffer to the correct
\sv{FULL} state.
What is really missing here is the buffer reference's ability to 
\emph{keep track of the dynamic change of its type/state} between
\sv{TypedRef[ProduceInt]} and \sv{TypedRef[ConsumeInt]}. 
Notice that statically we can only approximate these changes with a
common supertype, i.e. \sv{TypedRef[BufferInterf]}, but we have seen
that it is not enough. The alternative is to take a new reference, of
the suitable type, each time the state has changed.
This is what we will do in the next subsection resorting to the
\emph{Continuation-Passing} programming style.

It is important to observe that the combination of typed actor
references and continuation-passing style is exactly (part of) the
solution adopted by the Akka Typed \cite{AkkaTyped} experimental
library available in the latest Akka release. Further major
differences between our approach and that of Akka Typed will be
presented in the next sections.

\subsection{Tracking State Changes with Continuation-Passing Style}

In order to keep track of the state changes caused by message
reception, we let actors reply by sending to users a
\emph{continuation reference}, that is a reference to themselves
\emph{typed at the new state} that can be used to continue the
interaction according to the rest (the continuation) of the protocol.
Since in the actor model the communication is asynchronous,
messages are enriched with an additional parameter \sv{replyTo}
referring to the actor that expects the corresponding reply message 
carrying the continuation. 

Accordingly, in the code below the \sv{insert} and \sv{remove} case
classes have the additional parameter \sv{replyTo} of type
\sv{ActorRef} corresponding to the (not necessarily typed) user actor
that is waiting for the continuation. 
Moreover, the additional case classes defined in lines 6-7 stand for 
the reply messages sending back to the user the (typed)
continuations. More precisely, when the buffer receives an
\sv{insert} message it moves to the state \sv{FULL}, hence its
continuation, carried by the message \sv{insertReply}, has type
\sv{TypedRef[ConsumeInt]}. Similarly, the reply to a \sv{remove} request
carries both the removed value and a continuation of type
\sv{TypedRef[ProduceInt]} standing for the buffer that moved to the
state \sv{EMPTY}.

\begin{SI}
// parameters replyTo correspond to Buffer users, it is sufficient
// to type them with ActorRef since we do not model their protocol
case class insert[T](value:T, replyTo:ActorRef) extends ProduceInt
case class remove(replyTo:ActorRef) extends ConsumeInt
// reply messages carry the buffer's continuation IN THE NEW STATE
case class insertReply(o:TypedRef[ConsumeInt])
case class removeReply[T](v:T,o:TypedRef[ProduceInt])

class Buffer[T] extends Actor {
  def EMPTY:Receive = {
      case insert(x:T,r) => context.become(FULL(x))
                            r ! insertReply(new TypedRef[ConsumeInt](self))
  }
  def FULL(x:T):Receive = {
      case remove(r) => context.become(EMPTY)  
                        r ! removeReply(x, new TypedRef[ProduceInt](self))
  }
  def receive = EMPTY
}

val s = ActorSystem()
val untypedBuffer = s.actorOf(Props(new Buffer[Int]),"buff")
val buffer = new TypedRef[ProduceInt](untypedBuffer)

val user = s.actorOf(Props(new Actor{
    buffer tyTell insert(4,self) 
    
    def run(v:Int) :Receive = {
      case insertReply(o) =>  o tyTell remove(self)
                              //o tyTell insert(9,self) does not compile 
      case removeReply(x,o) =>  if (v > 0){
                                   o tyTell insert(v+4,self) 
                                   context.become(run(v-1))
                                } else  println("done")
    }   
    def receive = run(4)
  }))

\end{SI}
Compared to the previous implementation of the \sv{Buffer} actor, the
new definition takes care of the reply messages (lines 12 and 16).
The user actor undergoes major changes: the initial \sv{buffer}
reference has type \sv{TypedRef[ProduceInt]} (line 23) and represents
an initially empty buffer. This reference is used \emph{just once} by
the user in line 26 to ask for the insertion of the integer
\sv{4}. Subsequent references to the buffer are those bound by the
parameter \sv{o} of the reply messages (lines 29 and 31).  Therefore,
by rebinding the buffer object to new references as a result of
sending a message, and by typing such new references with a type that
describes the newly reached state, we can control that sequences of
message invocations conform to the buffer's usage
protocol, provided that each reference is used at most once.

For instance, sending in line 29 an \sv{insert} message to the
reference \sv{o} would not compile since the compiler knows that the
type of the variable \sv{o} is \sv{TypedRef[ConsumeInt]}. For the same
reason also the expression in line 30 would not compile, even if it
follows a \sv{remove} request.  On the other hand, two successive
calls \sv{o tyTell remove(self); o tyTell remove(self)} would be well
typed even if they do not respect the buffer protocol, and similarly
invoking \sv{tyTell insert(0,self)} on the reference \sv{buffer}
instead of \sv{o} in line 29 would not be prevented by the
compiler. The point is that in order to enforce protocol compliance,
it is important to ensure that a reference (to a typestate actor) is
used \emph{at most once}.

The behavioral type system developed in \cite{CrafaPadovani15}
statically checks the linear usage of typed object references, however
this is beyond the capabilities of the standard Scala type system.
Therefore, in order to enforce an affine usage of typed references,
that is at most one call of the \sv{tyTell} method, we could enrich
the \sv{TypedRef} class with an \emph{affine flag} and 
transform an affinity violation, i.e., a logical error, into a runtime
error. For the sake of completeness we show below the corresponding
definition, but we do not assume it in the rest of the paper. 

\begin{SI}
class TypedRef[-T](r:ActorRef) {
  private var used = false
  def tyTell(msg:T) = { 
    if (!used){ used = true; r ! msg }
    else {println("affinity violation"); throw new Exception{}}
  }
}
\end{SI}

We finally comment on the difference between this approach and that of
Akka Typed actors. As we said above, this module provides both typed
actor references (which are instances of the \sv{ActorRef[-T]} class)
and a continuation passing programming style. Type safety 
guarantees~\cite{BrusaferroCrafa16} 
that in well typed programs an actor of type
\sv{ActorRef[T]} will only receive messages of type \sv{T}. However, a
key assumption/constraint of Akka Typed actors is that they can
dynamically change their behavior only as along as the new behavior
still handles all the messages handled by the previous one.
For instance, the 1-place buffer should handle both \sv{insert} and
\sv{remove} messages in both states \sv{EMPTY} and \sv{FULL}.
In other words, the buffer must be prepared (i.e., programmed) to
receive messages that arrive when it is in the wrong state.
Here we are instead claiming that it is possible to define
typestate-oriented actors whose behaviors are defined only on messages
that are meaningful in the corresponding state. 

We will further discuss Akka Typed constraint in later sections; we
conclude observing that even in the Akka Typed module there is no
support to statically check whether the program violates the
continuation-passing style by re-using an ``out of date'' reference.
So, like in the code above, it is responsibility of the programmer to
use at most once a typed reference that is subject to a state/type
change. 

\subsection{Simplifying the Code with the Continuation Monad}

Adopting the continuation-passing style allows stateful actors to take
advantage of type checking, but it entails a more involved user
code. Explicit use of reply messages carrying continuations might be
useful to correctly combine the protocols of two (or more) interacting
stateful actors. However, in simple cases where a stateful actor is
accessed by a single user, it would be useful to simplify the user
code and bring to light its logic. Indeed, looking at the user code
shown in the previous subsection it is not immediate to understand
that the actor is alternating \sv{insert} and \sv{remove} requests.
 
In this section we illustrate how to simplify the code accessing a
stateful actor by encapsulating continuations into a monad.
%
We start by showing the refactoring of the code of a user that
alternates \sv{insert} and \sv{remove} requests to a 1-place buffer:

\begin{SI}
val user = s.actorOf(Props(new Actor{    
    for { o <- buffer tyTell insert(0)             
	  o <-  Buffer.afterInsert(o) tyTell remove() 
          o <-  Buffer.afterRemove(o) tyTell insert(2) 
          o <-  Buffer.afterInsert(o) tyTell remove() 
          o <-  Buffer.afterRemove(o) tyTell insert(4) 
          o <-  Buffer.afterInsert(o) tyTell remove()  
     } yield println(" END ")

    def receive =  PartialFunction.empty
}))
\end{SI}
This code is very similar to that in the untyped case, but it heavily
builds on typed continuations, which are hidden within the monad but
are still enforcing type safety.  Messages \sv{insertReply} and
\sv{removeReply} are not needed anymore, and the code adopts the
for-notation offered by Scala to work with monads. The monad we define
below makes use of futures so that the code above has the asynchronous
and non-blocking semantics that is distinctive of actors, and
the code style is reminiscent of Scala concurrent programs that use
futures. 

First of all, observe that by using the same variable \sv{o} for each
binding in lines 2-7, we reduce the risk (still not checked by the
compiler) of using an out-of-date reference. Similarly, it is easier
to spot whether the reference \sv{buffer} is used only once in line 2.
On the other hand, as we will explain below, the user code needs
explicit type-enforcements, that have been encapsulated in the methods
\sv{afterInsert} and \sv{afterRemove}. 

\noindent
The main ideas of this approach can be summarized as follows: 
\begin{enumerate}
\item we have already observed in Section 2.2 that it is essential
  for types to somehow encode information about the state changes
  (possibly) provoked by the reception of messages. In Section 2.3
  such information is encoded by the reply messages, e.g., the
  definition \sv{case class insertReply(o:TypedRef[ConsumeInt])} 
  links the \sv{insert} message with the \sv{ConsumeInt} interface
  assumed by the buffer in the next state \sv{FULL}. 
  Here we get rid of reply messages, and we rather define a
  \emph{usage protocol expression} that directly encodes the state 
  transitions involved by the actor protocol. More precisely, 
  a protocol expression is a function associating the next
  state, i.e. the typed continuation, to each message in the
  actor's interface (lines 27-38 below). 
\item Instead of \sv{TypedRef[T]} we use a richer actor wrapper
  \sv{ProtRef[T]}, which 
  encapsulates both a reference to a stateful actor \emph{at state}
  \sv{T} and the usage protocol expression of such actor.
  The actor's protocol is indeed inspected in the new definition of
  the \sv{tyTell} method, that now returns an object of type
  \sv{Continuation} which encapsulates the next state that will be
  assumed by the actor upon reception of the message passed as
  parameter to \sv{tyTell} (lines 13 and 15 below). 
\item The abstract class \sv{Continuation} has an abstract
  type member \sv{NextState} and encapsulates a promise and a paired
  future holding a value of type \sv{ProtRef[NextState]}.
  The pair (promise, future) allows the usual
  asynchronous and non-blocking semantics: the user of the
  actor sends a message and immediately obtains a future (as a
  continuation reference), that will be asynchronously completed by
  the actor which fulfills the corresponding promise upon handling
  the received message. 

  More precisely, when the user invokes \sv{tyTell(msg)}, the actor
  protocol is inspected and a pair (promise, future) of the suitable
  ``next type'' is created (see line 13 and the definition of the
  \sv{protocol} expression in lines 27-38 below). Then the promise is
  forwarded to the actor to be completed (line 14), while the
  continuation is returned to the caller, that is to the user (line
  15). Since the \sv{Continuation} class is implemented as
  a monad on top of the \sv{Future[T]} monad (lines 6-8), 
  the user can deal with the returned object using the simple
  for-notation that is common in Scala programming with 
  futures.
\end{enumerate}


\begin{SI}
abstract class Continuation {
  type T // T stands for the Next State
  val  p:Promise[ProtRef[T]]
  val  f:Future[ProtRef[T]]

  def map[S](fun: ProtRef[T] => S) :Future[S] = { f map fun }
  def flatMap[S](fun: ProtRef[T] => Future[S]) :Future[S] = { f flatMap fun }  
  def filter(pred: ProtRef[T] => Boolean):Future[ProtRef[T]] = { f filter pred }
}

class ProtRef[-T](r:ActorRef, protocol: T=>Continuation) {
  def tyTell(msg:T) :Continuation = { 
    val cont=protocol(msg) // returns a Promise of suitable type
    r!(msg,cont.p)         // r will complete the promise p
    return cont
  }
}

trait BufferInterf 
trait ProduceInt extends BufferInterf 
trait ConsumeInt extends BufferInterf  

case class insert(value:Int) extends ProduceInt // for simplicity just Int values
case class remove() extends ConsumeInt
 
object Buffer {  // protocol for a buffer whose user alternates insert and remove
 def protocol : BufferInterf => Continuation = {
   case insert(v) => new Continuation {
			type T= ConsumeInt
			val p=Promise[ProtRef[ConsumeInt]]
			val f=p.future
		      }
   case remove() =>  new Continuation {
                       type T=ProduceInt
                       val p=Promise[ProtRef[ProduceInt]]
                       val f=p.future
                      }
} 
 def afterInsert[U](o:ProtRef[U]):ProtRef[ConsumeInt] = o.asInstanceOf[ProtRef[ConsumeInt]]
 def afterRemove[U](o:ProtRef[U]):ProtRef[ProduceInt] = o.asInstanceOf[ProtRef[ProduceInt]]
}
\end{SI}

The definition of the Buffer actor must then be modified so that
instead of sending reply messages the message handler completes a
promise. Notice also that, because of the new definition of the
\sv{tyTell} method, the buffer's incoming messages are pairs made of a
message tag and the specific promise instance to be completed (lines 3
and 8 below).  The injection of the explicit type of the promise
variable \sv{p} in lines 3 and 8 is needed to typecheck the promise
completion in lines 5 and 10, on the other hand, a warning is produced
by the compiler since the type argument of the generic class
\sv{Promise} gets unchecked because it is eliminated by type
erasure. Anyway, the correct type parameter to inject in lines 3 and 8
can be automatically deduced from the buffer's protocol definition,
i.e. the \sv{Buffer.protocol} function defined in the companion object
(lines 27-38 of the code above).

\begin{SI}
class Buffer extends Actor {
  def EMPTY :Receive = {
     case (insert(x),p:Promise[ProtRef[ConsumeInt]]) => 
                                   context.become(FULL(x))
                                   p success new ProtRef[ConsumeInt](self,Buffer.protocol)
  }
  def FULL(x:Int) :Receive = {
      case (remove(),p:Promise[ProtRef[ProduceInt]]) =>
                                   context.become(EMPTY) 
                                   p success new ProtRef[ProduceInt](self,Buffer.protocol)
  }
  def receive = EMPTY
}
\end{SI}

For the sake of completeness, we show the program code that completes
the definition of the user actor shown at the beginning of this
section:

\begin{SI}
val s = ActorSystem()
val untypedBuffer = s.actorOf(Props(new Buffer),"buff")
val buffer = new ProtRef[ProduceInt](untypedBuffer,Buffer.protocol)
val user = ....as above... 
\end{SI}

\paragraph{Summary of the TSOP guarantees actually enforced by typing.}
The advantages of the approach we described in this section can be
summarized as follows:
\begin{itemize}
\item the \underline{buffer} is truly implemented as a type-stateful
  object: 
  ($i$) each state corresponds to an actor behavior that handles only
  the messages that belong to the interface assumed in that state, and
  ($ii$) state transition is achieved by changing the current behavior
  (by means of \sv{become}) and using a typed wrapper corresponding
  to the interface assumed in the new state;

\item the \underline{user} code has a clean logic and its compilation
  guarantees that the buffer will only receive {\bf the intended
    messages at the intended state}.  For instance, two consecutive
  \sv{remove} requests would result in the ill-typed expression
  \sv{Buffer.afterRemove(o) tyTell remove()} that is rejected by the
  compiler.

  It is worth observing that the type casts provided by the methods
  \sv{Buffer.afterInsert} and \sv{Buffer.afterRemove} are inevitable
  because the compiler can only statically assume that the
  continuation returned by an invocation of the \sv{tyTell} operator
  has an abstract type, therefore a type-coercion is needed.  The
  definition of the \sv{tyTell} method ensures that the actual type of
  the returned object is correct because it has been deduced from the
  usage protocol, hence no type-cast error will occur at runtime.
  Removing these casts requires implementing a specific static
  analysis; however our solution keeps the code clean by
  resorting to so-called Phantom Types\footnote{see for instance
    https://blog.codecentric.de/en/2016/02/phantom-types-scala/ or
    http://ktoso.github.io/scala-types-of-types/\#phantom-type}.
%
\item most of the code of the given program can be automatically
  generated from the specification of the intended buffer
  \underline{protocol}. In our code the protocol is made of 
  ($i$) a hierarchy of empty traits that encode interfaces,
  ($ii$) a set of case classes defining the messages belonging to
  these interfaces, and 
  ($iii$) a function expressing the state transitions entailed by
  message reception. Ideally, TSOP programmers could define object
  protocols using some other more declarative language, e.g. UML
  sequence diagrams, FSMs or CFSMs as in~\cite{LangeTuostoYoshida15}
  or even using a specific type language as in~\cite{CrafaPadovani15},
  and the translation into typed code analogous
  to the one above could be automatically generated. We leave this
  subject for future developments.
\end{itemize}


\subsection{A typestate-based buffer with multiple concurrent users}

So far we have considered a rather simple scenarion in which a
stateful actor is accessed by a single concurrent user. Ensuring that
a stateful object only receives the intended messages at the intended
states is much more difficult when the object is aliased or
shared~\cite{BierhoffAldrich07,DamianiEtAl08}.  For instance, if the
empty buffer is initially shared by two users and the first user fills
it with a value, it causes the buffer to move to the \sv{FULL} state
thus disabling the insertion that could have been concurrently
requested by the second user. However, if the first user later on
removes the inserted value, the concurrent insertion requested by the
second user could be only temporarily disabled, since the buffer
eventually gets back to the \sv{EMPTY} state.
 
In other terms, in a concurrent (and asynchronous) setting, an actor
must be prepared to receive messages that correctly belong to its
protocol even if the actor is not in the appropriate state \emph{yet}.
Dealing with this scenario is indeed the main reason why the Akka
Typed library allows an actor to dynamically change its behavior as
long as the new behavior accepts the same set of messages as the
original one. Indeed, if the buffer in the \sv{FULL} state can somehow
handle also the \sv{insert} messages, no requests get lost. Anyway, in
this way the stateful object essentially can change its behavior but
it always keeps the same interface.  We showed in
\cite{CrafaPadovani15} that we can instead keep the TSOP idea of
different interfaces in different states even in a concurrent setting
by resorting to the Chemical Semantics.

The chemical model of concurrency, due to Berry and Boudol in 1992
\cite{BerryBoudol92}, interprets the state of an asynchronous
concurrent system as a soup of molecules, corresponding to the
messages that have been sent to the concurrent entities. The behavior
of the system is then described by reaction rules, that specify how
(patterns of) molecules/messages can be consumed so to produce new
ones.  This model of concurrency is the foundation of the Join
Calculus~\cite{FournetGonthier96}, a formal language for which there
exist a number of both native and library implementations
(e.g.,~\cite{ItzsteinJasiunas03,Comega,Russo07,HallerVanCutsem08}).
On the other hand, a similar semantic model can be established also in
Scala Akka ($i$) by interpreting molecules as messages sent to actors,
($ii$) by interpreting actor behaviors as reaction rules, and ($iii$)
by modifying the default management of the actor mailbox so to keep a
received message/molecule until it triggers a reaction, i.e. until the
current behavior/state of the actor is able to handle it.

A precise formal account of the connection between the actor model and
the Join Calculus that allows to scale to actors the type safety
properties proved in \cite{CrafaPadovani15}  is subject of 
current work. Here we illustrate by means of examples that Scala Akka
actors can support the chemical semantics, and the
Scala type system can already check typestate-oriented code to some
extent. 

More precisely, we define a \sv{Chemical} trait that can be used to
optionally provide actors with chemical semantics by means of mix-in 
composition. The trait is defined so that it only keeps the messages
of (super)type \sv{ProtocolMsg}, that will eventually be handled since
they actually belong to the actor's protocol, while other messages can
be immediately discarded as errors. 

\begin{SI}
trait ProtocolMsg   // identifies the type of the messages to keep

trait Chemical extends Actor {
  private val soup :ArrayBuffer[ProtocolMsg] = new ArrayBuffer()
  private def check() = { soup.map(self ! _); soup.clear } 
  private def keep :Receive = { case msg:ProtocolMsg => soup.append(msg) } 
  
  def chemBecome(newState:Receive) = { check(); context.become(newState) }
  def chemReact(behave:Receive):Receive = behave orElse keep
}
\end{SI}
The \sv{chemReact} method extends the actor's behavior so to keep
protocol messages (i.e. messages of type \sv{ProtocolMsg}) arriving at
the wrong time, while \sv{chemBecome} is used to change the actor's
state and re-check saved messages. In this definition saved messages
are simply re-sent to the mailbox, we show in the next section a more
realistic implementation that relies on the Akka's \sv{Stash}
trait~\cite{StashURL} that allows an actor to temporarily stash away
messages and prepend them to the actor's mailbox right before changing
the current behavior.   

As far as our running example in concerned, we can obtain a 1-place
buffer that is robust with respect to multiple concurrent users by
mixing-in the \sv{Chemical} trait to its definition.  To illustrate,
we consider a buffer that can be accessed by two kinds of user actors:
a \sv{Producer} actor that repeatedly sends \sv{insert} messages and a
\sv{Consumer} actor that repeatedly sends \sv{remove} messages.  In
this scenario the buffer's usage protocol is different from the one we
considered in the previous sections: while the buffer still switches
between the states \sv{EMPTY} and \sv{FULL}, now each user is aware of
a single, fixed buffer interface, i.e., the \sv{Producer} relies just
on the \sv{ProduceInt} interface to send \sv{insert} messages, while
the \sv{Consumer} is aware just of the \sv{ConsumeInt} interface.
Accordingly, in the typed encoding of the new buffer's protocol, the
type of the continuation references must be adapted: after sending an
\sv{insert} message a \sv{Producer} will get a continuation of type
\sv{TypedRef[ProduceInt]}, while the \sv{Consumer} will obtain a
continuation of type \sv{TypedRef[ConsumeInt]} as a result of a
\sv{remove} message.  We show below the code corresponding to the
explicit passing of continuations (cf. section 2.3), while the
integration of the chemical semantics with the continuation monad is
postponed to the next section.

\begin{SI}
trait BufferInterf extends ProtocolMsg  // chemical molecules
trait ProduceInt extends BufferInterf
trait ConsumeInt extends BufferInterf  

case class insert[T](value:T, replyTo:ActorRef) extends ProduceInt
case class remove(replyTo:ActorRef) extends ConsumeInt
case class insertReply(o:TypedRef[ProduceInt])
case class removeReply[T](v:T,o:TypedRef[ConsumeInt])

class Buffer[T] extends Actor with Chemical {
  def EMPTY:Receive = chemReact {
      case insert(x:T,r) => chemBecome(FULL(x))		  
		            r ! insertReply(new TypedRef[ProduceInt](self))
   }
  def FULL(x:T):Receive = chemReact {
      case remove(r) => chemBecome(EMPTY)
			r ! removeReply(x, new TypedRef[ConsumeInt](self))
  }
  def receive = EMPTY
}

class Producer(buffer:TypedRef[ProduceInt]) extends Actor  {  
  buffer tyTell insert(0, self)  
  def sent(v:Int) : Receive = {
    case  insertReply(o) => o tyTell insert(v+10,self) //remove(self) would not compile
			    context.become(sent(v+10)) 
   }
  def receive = sent(0) 
}

class Consumer(buffer: TypedRef[ConsumeInt]) extends Actor { 
   buffer tyTell  remove(self)    
   def receive: Receive = {
    case  removeReply(x,o)=> o tyTell  remove(self)
   }
}

val s = ActorSystem()
val bufferUntyped = s.actorOf(Props(new Buffer[Int]),"buffer")
val buffer = new TypedRef[BufferInterf](bufferUntyped) //use contravariance of TypedRef[-T]
val producer1 = s.actorOf(Props(new Producer(buffer)))
val consumer1 = s.actorOf(Props(new Consumer(buffer)))
val consumer2 = s.actorOf(Props(new Consumer(buffer)))
\end{SI}

The program above spawns one producer and two consumers accessing the
same 1-place buffer. Because of nondeterminism it is unreasonable to
expect that the buffer \emph{receives} alternated messages \sv{insert}
and \sv{remove}. Thanks to its chemical semantics, however, it is
guaranteed that the buffer \emph{handles} the messages in the correct
sequence. Interestingly, the definition of the new class \sv{Buffer}
is identical to the previous ones but for the usage of \sv{chemReact}
and \sv{chemBecome} methods introduced by the \sv{Chemical}
mix-in. Therefore, we have that with a simple mix-in the clean TSOP
definition mentioning only the intended messages at the intended
states is actually robust in a fully concurrent context.

We finally remark that in lines 41-43 above we relied on the
contravariance of the type parameter of the class \sv{TypedRef[-T]}:
indeed, the \sv{buffer} reference, which is of type
\sv{TypedRef[BufferInterf]}, is passed as a parameter to the
\sv{Producer}, resp. \sv{Consumer}, constructor which instead
expects a parameter of super-type \sv{TypedRef[ProduceInt]},
resp. \sv{TypedRef[ConsumeInt]}.  


\subsubsection{Chemical semantics with the Continuation monad}

For the sake of completeness we rephrase the code above to show the
integration of the chemical semantics with the \sv{Continuation}
monad. In particular, the \sv{Chemical} trait is implemented by means
of the \sv{Stash} trait and it is modified (line 5) to keep the
specific kind of messages that in this scenario will be sent
to an \sv{Actor}, that is pairs \sv{(m,p)} where \sv{m} is the actual
message and \sv{p} is the promise instance to be filled upon message
handling.

The rest of the code is straightforward: the buffer protocol is
declared in the companion object and the \sv{Buffer} class mixes-in
the \sv{Chemical} trait and defines the actor in a typestate-oriented
style. Finally, the \sv{Producer} and \sv{Consumer} classes define
user actors by means of the for-notation.

\begin{SI}
trait ProtocolMsg

trait Chemical extends Actor with Stash {
  private def check() = { unstashAll()  }    
  private def keep:Receive = { case (m:ProtocolMsg,p) => stash() }
			       
  def chemBecome(newState:Receive) = { check(); context.become(newState) }
  def chemReact(behave:Receive):Receive = behave orElse keep
}

object Buffer {  // protocol for a buffer with concurrent producers and consumers
 def protocol : BufferInterf => Continuation = {
   case insert(v) => new Continuation {
                        type T= ProduceInt
                        val p=Promise[ProtRef[ProduceInt]]
                        val f=p.future
                      }
   case remove() =>  new Continuation {
                       type T=ConsumeInt
                       val p=Promise[ProtRef[ConsumeInt]]
                       val f=p.future
                      }
}
 def afterInsert[U](o:ProtRef[U]):ProtRef[ProduceInt] = o.asInstanceOf[ProtRef[ProduceInt]]
 def afterRemove[U](o:ProtRef[U]):ProtRef[ConsumeInt] = o.asInstanceOf[ProtRef[ConsumeInt]]
}

class Buffer extends Actor with Chemical{
    def EMPTY :Receive = chemReact {
      case (insert(x),p:Promise[ProtRef[ProduceInt]]) => 
		    chemBecome(FULL(x))
	            p success new ProtRef[ProduceInt](self,Buffer.protocol)
    }
    def FULL(x:Int) : Receive = chemReact { 
      case (remove(),p:Promise[ProtRef[ConsumeInt]]) => 
		    chemBecome(EMPTY)
                    p success new ProtRef[ConsumeInt](self,Buffer.protocol)
    }
    def receive = EMPTY
}

class Producer(buffer:ProtRef[ProduceInt]) extends Actor { 
   for { o <- buffer tyTell insert(0)
         o <- Buffer.afterInsert(o) tyTell insert(10)
         o <- Buffer.afterInsert(o) tyTell insert(20)
         o <- Buffer.afterInsert(o) tyTell insert(30)
         o <- Buffer.afterInsert(o) tyTell insert(40)
   } yield println("End Producer")

  def receive = PartialFunction.empty 
}

class Consumer(buffer:ProtRef[ConsumeInt]) extends Actor{
  for { o <- buffer tyTell remove()
        o <- Buffer.afterRemove(o) tyTell remove()
        o <- Buffer.afterRemove(o) tyTell remove()
        o <- Buffer.afterRemove(o) tyTell remove()
   } yield println("End Consumer")

  def receive = PartialFunction.empty 
}

val s = ActorSystem()
val bufferUntyped = s.actorOf(Props(new Buffer),"buffer")
val buffer = new ProtRef[BufferInterf](bufferUntyped,Buffer.protocol) //use contravariance
val producer = s.actorOf(Props(new Producer(buffer)))
val consumer = s.actorOf(Props(new Consumer(buffer)))
\end{SI}

\section{Bookshop Server}

In this section we address a more complex example, showing that our
solution for typestate-oriented actor programming is effective even
with more complex protocols. More precisely, we consider an example
coming from the literature about session types, i.e. the Bookshop
Server (cf. \cite{GayVasconcelos10}). A client accessing the server
initially adds a number of items to the shopping basket, then
asks for the checkout and sends information about the credit card and
the shipping. 
The bookshop protocol is defined by the recursive session type
\sv{Shop = \&<add:?Book.Shop, checkout:?Card.?Address.end>}, which
essentially corresponds to the TSOP protocol defined by the following
FSM: 
\begin{center}
\includegraphics[width=12cm]{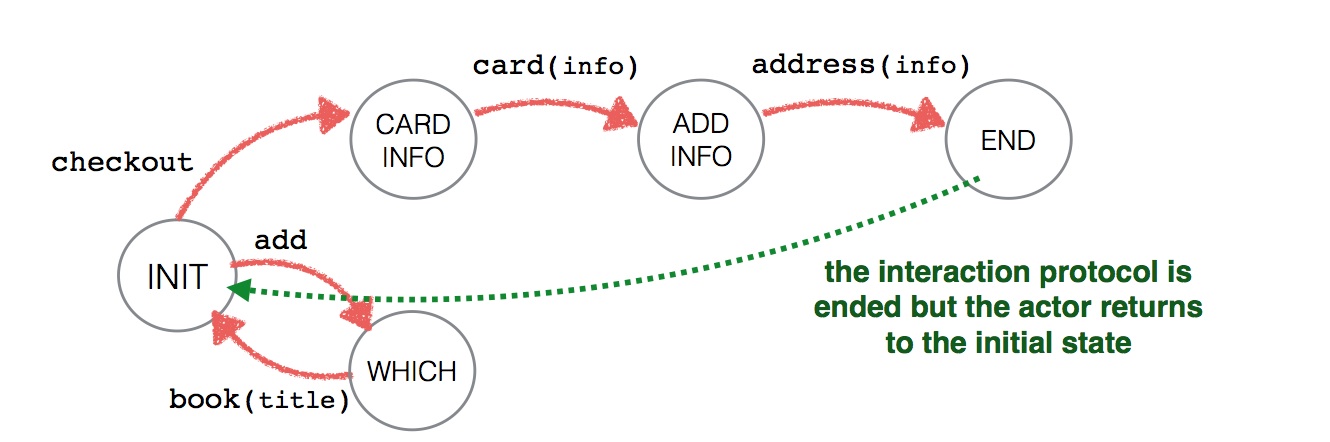}
\end{center}

In order to define this protocol in terms of Scala types, we first
choose the names of the five possible states and the corresponding
interfaces (lines 2-6 and 8-12). State transitions are then encoded in
the \sv{Shop} companion object (lines 15-41). 
Since the bookshop can be accessed by multiple concurrent users, we
need to decide a serving policy. Initially we let the bookshop
completely serve one user at a time, and we rely on the chemical
semantics to keep initial requests that may concurrently arrive from a
user while serving another one. Therefore, we tag with type
\sv{ProtocolMsg} only the initial message \sv{add} (line 8).
We will later discuss other choices that allow a greater degree of
concurrency. 

\begin{SI}
trait ShopInterface
trait InitInterf extends ShopInterface  // with ProtocolMsg
trait WhichInterf extends ShopInterface
trait CInfoInterf extends ShopInterface
trait AddInfoInterf extends ShopInterface
trait EndInterf extends ShopInterface

case class add(name:String) extends InitInterf with ProtocolMsg
case class checkout(name:String) extends InitInterf
case class book(name:String,b:String) extends WhichInterf
case class card(name:String,cardNum:String) extends CInfoInterf
case class address(name:String,add:String) extends AddInfoInterf

object Shop {
 def protocol :ShopInterface => Continuation = {  
   case add(n) => new Continuation {
		     type T=WhichInterf
		     val p=Promise[ProtRef[WhichInterf]]
                     val f=p.future
                  }
   case checkout(n) => new Continuation {
			 type T= CInfoInterf
			 val p=Promise[ProtRef[CInfoInterf]]
                         val f=p.future
                       }
   case book(n,b) => new Continuation {
		       type T=InitInterf	
		       val p=Promise[ProtRef[InitInterf]]
		       val f=p.future
                     }
   case card(n,cn) => new Continuation {
		       type T=AddInfoInterf
		       val p = Promise[ProtRef[AddInfoInterf]]
                       val f=p.future
                       }
   case address(n,add) => new Continuation {
			   type T=EndInterf 
			   val p=Promise[ProtRef[EndInterf]]
			   val f=p.future
                          }
  }
  def afterAdd[U](o:ProtRef[U]):ProtRef[WhichInterf] = o.asInstanceOf[ProtRef[WhichInterf]]
  def afterCo[U](o:ProtRef[U]):ProtRef[CInfoInterf] = o.asInstanceOf[ProtRef[CInfoInterf]]
  def afterBook[U](o:ProtRef[U]):ProtRef[InitInterf] = o.asInstanceOf[ProtRef[InitInterf]]
  def afterCard[U](o:ProtRef[U]):ProtRef[AddInfoInterf]=o.asInstanceOf[ProtRef[AddInfoInterf]]
  def afterAddress[U](o:ProtRef[U]):ProtRef[EndInterf] = o.asInstanceOf[ProtRef[EndInterf]] 
}

class Shop extends Actor with Chemical {
  private val shopBasket:Map[String,String] = Map[String,String]()

  def INIT:Receive = chemReact {
    case (add(n),p:Promise[ProtRef[WhichInterf]]) => println(n+" please chose a book")
		                p success new ProtRef[WhichInterf](self,Shop.protocol)
		                context.become(WHICH)
		    
    case (checkout(n),p:Promise[ProtRef[CInfoInterf]]) => println("start payment for "+n)
		                p success new ProtRef[CInfoInterf](self,Shop.protocol)
		                context.become(CINFO)
  }

  def WHICH : Receive = chemReact {
    case (book(n,b),p:Promise[ProtRef[InitInterf]]) => println(b+"put in the basket of "+n)
		                if(shopBasket.contains(n))  shopBasket(n) += (" "+b)
                                else  shopBasket += (n->b)
                                p success  new ProtRef[InitInterf](self,Shop.protocol)
                                context.become(INIT) // ensures that the current user 
                                // completes the shopping without interleaving other users
                      // or use chemBecome(INIT)  to RECHECK SOUP!
  }

  def CINFO :Receive = chemReact {
    case (card(n,c),p:Promise[ProtRef[AddInfoInterf]]) => println("using card n."+c+"of "+n)
                                p success new ProtRef[AddInfoInterf](self,Shop.protocol)
		                context.become(ADDINFO)
  }

  def ADDINFO :Receive = chemReact {
    case (address(n,a),p:Promise[ProtRef[EndInterf]])=> 
                                println("shipping "+shopBasket(n)+" to "+n+" in "+a) 
	                        shopBasket.remove(n)
		                p success new ProtRef[EndInterf](self,Shop.protocol)
		                chemBecome(INIT) // rechecks the soup searching for
                               // another client to serve 
  }

  def receive = INIT
}
\end{SI}
Notice that the \sv{Shop} actor is defined as a stateful object with
four different behaviors, each one handling only the messages expected
to be received in the corresponding state. The additional state
\sv{END} appearing in the public protocol is only used by the users:
indeed, at the end of the interaction, the users receive a
continuation of type \sv{ProtRef[EndInterf]} (line 82) that ensures
that no further message can be sent to the \sv{Shop} actor since
\sv{EndInterf} corresponds to the shop's empty interface.

Every shop's behavior calls the \sv{chemReact} method 
to keep \sv{add} messages that might arrive at any moment from a user
different form the one that is currently served. On the other
hand, the \sv{chemBecome} method is called only in line 83 at the end
of the interaction with the current user. Indeed, while
\sv{context.become} simply changes the actor's state/behavior, the
\sv{chemBecome} method also re-sends the saved \sv{add} messages so
that they can be eventually handled.

In order to increase the throughput of the server, we can let the
bookshop interleave the shopping phases and the checkout phases of 
different users. More precisely, each time the shop gets back to state
\sv{INIT}, it can start serving the shopping or the checkout of
another actor. To change the policy in this way it is sufficient
to ($i$) collect in the chemical soup both the \sv{add} and
\sv{checkout} messages, that is to tag with type \sv{ProtcolMsg} the
entire \sv{InitInterf} interface (line 2), and ($ii$) allows a recheck
of the saved messages each time the shop enters the \sv{INIT} state,
which means using \sv{chemBecome} instead of \sv{context.become} in
line 67.


We conclude by showing the code of a program that spawns a bookshop
and three users that buy a couple of books each: 

\begin{SI}
class User(shop:ProtRef[InitInterf], name:String,info:Map[String,String]) extends Actor {
   for {
     o <- shop tyTell add(name) 
     o <- Shop.afterAdd(o)  tyTell  book(name,info("book1"))       
     o <- Shop.afterBook(o) tyTell  add(name) 
     o <- Shop.afterAdd(o)  tyTell  book(name,info("book2"))
     o <- Shop.afterBook(o) tyTell  checkout(name)
     o <- Shop.afterCo(o)   tyTell  card(name,info("cc"))
         // Shop.afterCo(o)  tyTell address(...)  shipping without paying does not compile!
     o <- Shop.afterCard(o) tyTell address(name,info("addr"))
   } yield println(name+" ended shopping")
 
  def receive=PartialFunction.empty
}

val s = ActorSystem()
val untypedShop = s.actorOf(Props(new Shop),"shop")
val shop = new ProtRef[InitInterf](untypedShop,Shop.protocol)

val infoM1=Map("book1"->"Pride and Prejudice","book2"->"Odissea","cc" ->"1234","addr"->"Padua")
val infoM2=Map("book1"-> "Ben Hur", "book2"->"Pinocchio","cc" ->"1212", "addr" -> "Venice")
val infoM3=Map( "book1"-> "Java8", "book2"->"Scala","cc" ->"8888", "addr" -> "NewYork")
val user1 = s.actorOf(Props(new User(shop,"Mary",infoM1)))
val user2 = s.actorOf(Props(new User(shop,"Jane",infoM2)))
val user3 = s.actorOf(Props(new User(shop,"Alice",infoM3)))
\end{SI}

Notice that in this code any attempt at sending an \sv{address}
message just after the message \sv{checkout} without sending the
credit card number, results in a type safety error. Indeed, the
reference \sv{Shop.afterCo(o)} in line 8 has type 
\sv{ProtRef[CInfoInterf]}, hence it only accepts \sv{card} messages.
On the other hand, this typing can only prevent wrong message sends
but cannot guarantee that intended messages will be actually sent. For
instance, a user which starts the interaction with the bookshop but
never sends the final \sv{address} message would satisfy the compiler
even if it would leave the shop in an intermediate state. The full
behavioral typing we studied in \cite{CrafaPadovani15} captures
these issues byt it requires linear types, which are not available in
the standard Scala compiler.


\section{Conclusions}

The aim of the TSOP approach is opposite to that of defensive
programming: instead of adding code to the stateful object to protect
it from wrong accesses, it checks the user code and statically marks
the wrong accesses as untyped actions.  We illustrated by means of
examples that Scala Akka effectively supports typestate-oriented actor
programming. An actor can be defined as a stateful object by letting
states correspond to actor behaviors that possibly handle different
sets of messages. Then, a dynamic change of behavior corresponds to a
change of actor's interface. Moreover, by fully encoding stateful
protocols into typed expressions, we can let the Scala compiler check
that the actor will only handle intended messages at the intended
times.  Nevertheless, a fair amount of defensive programming is still
needed since static detection of violations of protocol obligations
(e.g., ensuring that an intended message is eventually sent) requires
the verification of linear properties that are currenlty beyond the
expressive power of the Scala type system. Moreover, the integration
of TSOP with the fault tolerance properties required in distributed
actor systems is still an open research issue.

We conclude observing that even if TSOP enables the coordination of
concurrent entities, it certainly does not solve the general
problem. Indeed a protocol-oriented programming paradigm requires more
than a single conceptual model and suitable abstractions and
programming patterns can only emerge as a productive integration of
different techniques, both at the high level (i.e, techniques to
express and verify coordination properties) and at the
lower level (i.e., language abstractions, performace and scalability
issues). 
   

%

\section*{Ackwoledgments}
We thank the passionate people of Scala by the Lagoon, the Venice area
Scala user group, for the insightful comments and the stimulating
discussions. 


\end{document}